\documentclass[11pt]{article}

\usepackage[preprint]{acl}

\usepackage{times}
\usepackage{latexsym}
\usepackage[T1]{fontenc}
\usepackage[utf8]{inputenc}
\usepackage{microtype}
\usepackage{inconsolata}
\usepackage{graphicx}
\usepackage{orcidlink}

\usepackage{amsmath}
\usepackage{amssymb}
\usepackage{bm}
\usepackage{booktabs}

\newcommand{\bxi}{\bm{\xi}}
\newcommand{\bbeta}{\bm{\beta}}
\newcommand{\bSig}{\bm{\Sigma}}
\newcommand{\bLam}{\bm{\Lambda}}
\newcommand{\bOmega}{\bm{\Omega}}
\newcommand{\bmu}{\bm{\mu}}
\newcommand{\bb}{\mathbf{b}}
\newcommand{\bV}{\mathbf{V}}
\newcommand{\bI}{\mathbf{I}}
\newcommand{\bX}{\mathbf{X}}
\newcommand{\bmm}{\mathbf{m}}
\newcommand{\Norm}{\mathcal{N}}
\newcommand{\rmc}{{\mathrm{c}}}
\newcommand{\refsc}{{\mathrm{ref}}}

\title{Bayesian Consensus Calibration of Continuously Evolving IRT Item Banks}

\author{%
  Paul A. Jewsbury\orcidlink{0000-0001-5571-4623} \and Steven W. Nydick\orcidlink{0000-0002-2908-1188} \and Manqian Liao\orcidlink{0000-0002-8444-9440} \and Siyuan (Marco) Chen\orcidlink{0000-0002-3346-5424} \\
  Duolingo \\
  \texttt{\{paul.jewsbury, steven, mancy, marco\}@duolingo.com}}

\newcommand\blfootnote[1]{
  \begingroup
  \renewcommand\thefootnote{}\footnote{#1}
  \addtocounter{footnote}{-1}
  \endgroup
}

\begin{document}
\maketitle

\blfootnote{Accepted at the Artificial Intelligence in Measurement and Education Conference (AIME-Con), October 2026. To appear in \emph{Proceedings of AIME-Con}, National Council on Measurement in Education.}

\begin{abstract}
AI-based item generation and NLP-based prediction of item parameters are producing item banks that are substantially larger, sparser, and more frequently updated than conventional banks. Hierarchical Bayesian item response theory (IRT) is a natural calibration framework for such banks, but the common practice of refitting the entire accumulated response history at each update is costly and can exceed available memory. We describe \emph{consensus calibration}, a divide-and-conquer procedure that calibrates each time period independently and reconstructs the pooled posterior in two layers. First, the posterior draws of each period are mapped to a common metric by a robust characteristic-curve linking (Haebara) that is solved separately for each draw, which propagates the uncertainty of the linking transformation into the linked posteriors. Second, the linked item posteriors are combined as a product of Gaussian densities from which the population prior contributed by each period is removed and a single prior---obtained by consensus across the per-period population posteriors---is reinstated. The correction targets the posterior dispersion, not only its location. As evidence for consensus calibration, we compare it to a pooled single-run analysis on a large operational assessment in terms of item-parameter recovery, an uncertainty-by-exposure diagnostic, and the ability distributions.
\end{abstract}

\section{Introduction}
\label{sec:intro}

Assessment programs increasingly generate items automatically and assign their parameters by prediction from item content rather than solely by pretesting. The Duolingo English Test, for instance, combines automatic generation, feature-based parameterization, and frequent recalibration of a continuously evolving bank (\hyperlink{cite.vondavier2024itemfactory}{A.~A. von Davier}, \citeyear{vondavier2024itemfactory}; \citealp{nydick2026maintaining}). The resulting response data are large, sparse, and adaptively sampled. Reproducing what a single fit to all of this accumulated data would yield, but without refitting past data as the bank grows, is the problem this paper addresses.

Explanatory hierarchical Bayesian IRT models suit this setting: the prior on each item's parameters is an estimated population distribution conditional on item features \citep{deboeck2004explanatory}. Sparsely observed items are shrunk toward their feature-based predictions, and the posterior supplies the parameter uncertainty on which adaptive selection and scoring depend \citep{fox2010bayesian,sharpnack2026}. The operational difficulty lies in recalibration: standard practice refits a single model to all accumulated data at each update, so cost grows continuously and, at contemporary bank sizes, holding the full history in memory may exceed available computational resources \citep{nydick2026spice}.

We treat recalibration as a divide-and-conquer Bayesian computation \citep{xu2024divide}. Each period is calibrated independently, and the period posteriors are combined without revisiting earlier data. Two features distinguish our case from the standard divide-and-conquer approach. First, each period's fit identifies the latent metric only up to a linear transformation, so the period posteriors are not directly comparable and must be linked. Second, the prior is not fixed but is itself estimated and hierarchical: each period's posterior incorporates its own population prior, which the combination must replace with a single prior to avoid understating uncertainty.

This paper makes four contributions: (i)~a per-draw robust linking step that maps each period to a common metric, adopting the sampling strategy of \citet{baldwin2011common} with a response-count-weighted robust characteristic-curve criterion across multiple periods, so that the uncertainty of the linking transformation is propagated into the period posteriors rather than treated as a fixed transformation with an added error term; (ii)~a two-layer consensus aggregation that combines the linked item posteriors and replaces the prior contributed by each period with a single prior obtained by consensus across the per-period population posteriors; (iii)~an explicit identifiability condition with a population-prediction fallback for items not identified by the period data beyond the population model; and (iv)~an empirical evaluation that compares consensus calibration with a pooled benchmark, illustrated on a large operational assessment.

\section{Background and Related Work}
\label{sec:related}

Automatic item generation increasingly relies on large language models, resulting in item banks much larger than was possible with manual authoring (\citealp{gierl2013aig}; \hyperlink{cite.vondavier2018aig}{M.~von Davier}, \citeyear{vondavier2018aig}; \hyperlink{cite.vondavier2024itemfactory}{A.~A. von Davier}, \citeyear{vondavier2024itemfactory}). Independently, NLP models predict item difficulty and related parameters from item content---using engineered text features, pretrained-language-model embeddings, or automated machine learning over both---so that items can be operationalized with little or no pretesting \citep{settles2020machine,mccarthy2021jumpstarting,yancey2024bertirt,sharpnack2024autoirt}. \citet{benedetto2023survey} and \citet{alkhuzaey2024review} survey text-based difficulty estimation, which centers on language assessment but reaches content-knowledge domains such as computer science and medicine. A recent shared task, for instance, benchmarked such systems on high-stakes medical items \citep{yaneva2024bea}. These methods are deployed operationally, where feature-based parameters let newly generated items enter scoring before substantial response data accrue (\hyperlink{cite.vondavier2024itemfactory}{A.~A. von Davier}, \citeyear{vondavier2024itemfactory}; \citealp{nydick2026maintaining,nydick2026spice}).

In hierarchical Bayesian IRT, item parameters are drawn from a population distribution whose hyperparameters are estimated jointly with the items \citep{fox2010bayesian}. Explanatory IRT generalizes the population model to a regression of item parameters on item features \citep{deboeck2004explanatory}. This regression is the object that feature-based parameter prediction estimates and that Bayesian calibration engines implement at scale (e.g., \citealp{nydick2026spice}), so the prior in such a calibration is the feature-conditional distribution of the item parameters.

Placing separately calibrated forms on a common metric is the problem of linking and equating, commonly solved by characteristic-curve methods that estimate a linear transformation by matching model-implied curves on common items \citep{haebara1980,stockinglord1983,kolen2014equating}. Exact invariance of the linking items across calibrations is neither realistic nor a prerequisite for valid comparison \citep{robitzsch2023invariance}. Characteristic-curve criteria fit the transformation to the anchors in aggregate, and robust variants attenuate the influence of the minority of items that function differently \citep{robitzsch2024sirt}, suiting them to this partial-invariance context. Because the link is itself estimated from the same data as the item parameters, linking error is generally dependent on sampling and measurement error. \citet{jewsbury2025variance} proposes and validates a taxonomy of variance-estimation methods that propagate linking error while accounting for these dependencies \citep{jewsbury2025invariance,jewsbury2026achievement}. \citet{baldwin2011common} carries this error in the posterior sample itself: the transformation is estimated from a single draw of each calibration and applied to that draw, repeated until the transformed posterior is assembled. We adopt this strategy for the multi-period case. This complements Bayesian nonparametric equating, which links observed score distributions \citep{karabatsos2009bayesian}, by performing IRT-parameter linking in a Bayesian setting.

Combining posteriors computed on disjoint subsets of a single dataset is the subject of consensus Monte Carlo \citep{scott2016consensus} and embarrassingly parallel MCMC \citep{neiswanger2014embarrassingly}, which parallelize one fit by partitioning the data across workers and merging the subset posteriors as a product of densities \citep{hinton2002poe,tresp2000bcm}. For IRT, \citet{xu2024divide} apply this within a single calibration, randomly partitioning the examinee sample and merging the subset posteriors by a Wasserstein barycenter, with the identification constraints held fixed so that no linking is required.

Periodic recalibration is a different problem. Applied to an evolving bank, these single-fit methods would re-analyze all the data at each update, treating it as one dataset to be partitioned and fit together rather than a sequence of runs to be combined. The proposed method instead calibrates each period once and combines a new period with the already-computed posteriors of the earlier runs, so an update's cost scales with the new data rather than the full history. This difference forces two departures from the single-calibration setting: (a)~because the periods are fit separately rather than within one run, they do not share a latent metric and must be linked; and (b)~because each period is its own hierarchical fit, it carries an estimated, period-specific prior that the combination must reconcile to a single prior. Consensus calibration addresses both.

\section{Setting and Notation}
\label{sec:setting}

Let $m=1,\dots,M$ index time periods, each calibrated independently by a Bayesian explanatory IRT engine \citep{nydick2026spice}. The calibration of period $m$ returns posterior draws of each item's parameter vector $\bxi_j$ and of the population hyperparameters of each item group: the coefficients $\bbeta$ of a multivariate regression of item parameters on item features and a residual covariance $\bSig$, so that the population model is $\bxi_j\sim\Norm(\bX_j\bbeta,\bSig)$ given features $\bX_j$. Each item is assumed to load on a single latent dimension, which fixes the orientation of the latent space; item responses then determine each dimension only up to its origin and unit, so the period-$m$ metric is identified only up to a linear transformation,
\begin{equation}
\theta_{\refsc} = A_m\,\theta + B_m, \qquad A_m>0,
\label{eq:scale}
\end{equation}
with constants specific to the period and dimension (the dimension index is suppressed). A designated reference period has $(A_m,B_m)=(1,0)$. The objective is to reconstruct the posterior that a single analysis pooling all periods (i.e., the full dataset) would yield.

The method uses only two properties of the item parameterization. First, each item $j$ has an expected-score (characteristic) curve $E_j(\theta)$, the expected item score at latent value $\theta$; matching these curves on anchor items is what estimates the linking constants (Section~\ref{sec:linking}). Second, the change of metric \eqref{eq:scale} induces a transformation of $\bxi_j$; applying the transformation draw-wise is what carries every item's posterior onto the reference metric.

\paragraph{Assumptions.}
The procedure relies on four assumptions. (i)~\emph{Disjoint examinee samples}, so that the period likelihoods are conditionally independent and the pooled likelihood factors across periods (Section~\ref{sec:consensus}). (ii)~\emph{Simple structure}: each item loads on a single latent dimension, reducing the metric indeterminacy to the linear map~\eqref{eq:scale}. (iii)~\emph{Aggregate anchor invariance}: the metric change is captured by $(A_m,B_m)$ and the anchor set is invariant in aggregate (Section~\ref{sec:linking}). (iv)~\emph{Gaussian summaries}: each period's item posterior, prior contribution, and population-coefficient posterior are summarized by Gaussian moments (Appendix~\ref{app:deriv}).

\section{Method}
\label{sec:method}

We aggregate independently calibrated periods rather than updating sequentially---carrying each period's posterior forward as the next period's prior---for three reasons. First, fitting each period as a self-contained hierarchical model lets its own population prior stabilize its sparsely observed items, rather than a prior whose form was fixed in advance. Second, the model specification, including the item parameter prediction model, may change between periods. Third, aligning the periods needs weaker invariance than sequential updating: its reused item priors presume each item is invariant, whereas a transformation fit to the anchors collectively needs only the anchor set to hold in aggregate \citep{robitzsch2023invariance}.

\subsection{Per-draw robust linking}
\label{sec:linking}

Under \eqref{eq:scale}, the expected-score curve of item $j$ expressed on the reference metric is
\begin{equation}
E_j^{m\to\refsc}(\theta) = E_j\!\big((\theta-B_m)/A_m\big),
\label{eq:curvemap}
\end{equation}
so the constants $(A_m,B_m)$ determine a transformation of the item parameters (for example, locations map as $d\mapsto A_m d+B_m$ and log-discriminations as $\log a\mapsto \log a-\log A_m$). The linking constants for period $m$ are chosen to minimize the weighted discrepancy between the reference curves and the transformed period-$m$ curves over an anchor set $\mathcal{A}$. Writing the discrepancy at $\theta$ as $D_j(\theta)=E_j^{\refsc}(\theta)-E_j^{m\to\refsc}(\theta)$,
\begin{equation}
H_m(A,B) = \int \sum_{j\in\mathcal{A}} \omega_j\,\rho\big(D_j(\theta)\big)\, w(\theta)\, d\theta,
\label{eq:haebara}
\end{equation}
where $w$ is a weight on the latent metric and $\omega_j$ is proportional to the number of responses on which anchor $j$ was calibrated. Weighting anchors by their response count lets better-estimated items contribute more to the link, approximating concurrent (joint) calibration, in which items with more data exert more influence on the common metric. The integral is evaluated by quadrature. The loss $\rho$ is a pseudo-Huber loss function, a smooth surrogate for the least-absolute ($L_1$) criterion of robust Haebara linking \citep{robitzsch2024sirt}. Note that any of the moment or characteristic-curve linking methods unified by \hyperlink{cite.vondavier2007unified}{M.~von Davier and A.~A. von Davier} (\citeyear{vondavier2007unified}) could replace Haebara here.

The criterion \eqref{eq:haebara} is minimized separately for each posterior draw, in the manner of \citet{baldwin2011common}, rather than once at the posterior mean. For draw $s$, the anchor curves are evaluated at that draw's parameter values and \eqref{eq:haebara} is solved to obtain $(A_m^{(s)},B_m^{(s)})$; that draw's transformation is then applied to draw $s$ of every item, anchors and non-anchors alike. The collection $\{(A_m^{(s)},B_m^{(s)})\}_{s=1}^{S}$ is a sample from the posterior of the linking constants. Applying it draw-wise propagates the linking uncertainty, together with its dependence on the item parameters, into the linked posteriors. Within the posterior, draw-wise linking plays the role that dependence-aware variance estimation plays for frequentist linking, where treating linking error as independent of the other sources of error misstates the uncertainty of downstream comparisons \citep{jewsbury2025variance,jewsbury2026achievement}.

Two filters are applied to the anchor set $\mathcal{A}$ before linking. The first removes any anchor whose potential scale reduction factor $\widehat{R}$ \citep{gelman1992rhat}, maximized over the item's parameters, exceeds $1.05$ in either the period being linked or the reference period. The second is an iterative purification \citep{candell1988iterative}: within each dimension, a provisional mean--sigma link \citep{kolen2014equating} is fit to the anchors' trait-metric locations (posterior means), anchors whose absolute residuals from the provisional link fall in the top decile are removed, and the link is refit, for at most five iterations. Here the reference period denotes the metric-defining calibration, not the pooled benchmark of Section~\ref{sec:eval}.

\subsection{Consensus aggregation and prior de-duplication}
\label{sec:consensus}

Linking places each item's period posteriors on the reference metric; aggregation now combines them into the pooled posterior, first as a precision-weighted product and then with a correction for the prior that each period contributes. Consider one item, with its index suppressed, and let its linked posterior in period $m$ be summarized by a Gaussian with mean $\hat\bmu_m$ and precision $\bLam_m=\hat\bSig_m^{-1}$. Sums over $m$ run over the periods in which the item was observed. If the periods comprise disjoint examinee samples, the joint likelihood factors across periods given the item parameters, and the pooled posterior is proportional to the product of the period posteriors with the multiply-counted prior removed. For the Gaussian summaries the product is Gaussian, with
\begin{equation}
\tilde\bLam = \sum_m \bLam_m, \qquad
\tilde\bmu = \tilde\bLam^{-1}\sum_m \bLam_m\hat\bmu_m .
\label{eq:product}
\end{equation}

Each factor in \eqref{eq:product} incorporates the population prior, so the product incorporates it $M$ times and is over-concentrated. Let the prior contributed by period $m$ be Gaussian with precision $\bOmega_m$ and mean $\bmm_m$---the moments of that period's population prediction for the item, mapped to the reference metric by the same draw-wise transformation---and let the single replacement prior have precision $\bOmega^\star$ and mean $\bmm^\star$. The corrected natural parameters are
\begin{equation}
\begin{aligned}
\bLam_\rmc &= \sum_m \bLam_m - \sum_m \bOmega_m + \bOmega^\star,\\
\bb_\rmc &= \sum_m \bLam_m\hat\bmu_m - \sum_m \bOmega_m\bmm_m + \bOmega^\star\bmm^\star,
\end{aligned}
\label{eq:Lc}
\end{equation}
with consensus mean $\bmu_\rmc=\bLam_\rmc^{-1}\bb_\rmc$ (Appendix~\ref{app:deriv}). When the periods share a fixed prior, \eqref{eq:Lc} reduces to $\bLam_\rmc=\sum_m\bLam_m-(M-1)\bOmega$, the combination rule of the Bayesian committee machine \citep{tresp2000bcm}. The general form of \eqref{eq:Lc} is required because the prior is estimated and differs across periods.

The replacement prior is obtained by consensus at the population layer. Let the period-$m$ posterior of the population coefficients $\bbeta$, mapped to the reference metric, be approximately Gaussian with mean $\bb_m$ and covariance $\bV_m$. Under a weak zero-mean Gaussian hyperprior with precision $\tau$, the consensus coefficients are the Gaussian product of the per-period posteriors with the hyperprior counted once,
\begin{equation}
\bbeta^\star = \Big(\sum_m \bV_m^{-1}-(M-1)\tau\bI\Big)^{-1}\sum_m \bV_m^{-1}\bb_m .
\label{eq:betastar}
\end{equation}
The consensus residual covariance $\bSig^\star$ is the item-count-weighted mean of the per-period residual covariances taken in log-Cholesky coordinates---the Fr\'{e}chet mean under the log-Cholesky metric \citep{lin2019cholesky,pinheiro1996unconstrained}---which is positive definite by construction (Appendix~\ref{app:logchol}). The replacement prior of \eqref{eq:Lc} for the item under consideration is then its consensus population prediction: $\bmm^\star=\bX\bbeta^\star$, with $\bX$ its features, and $\bOmega^\star=(\bSig^\star)^{-1}$. The population layer presupposes a common population-model specification across periods, as in our application; under specification changes, the replacement prior can instead be formed from the item-level predictions. Consensus thus applies at both layers---the item layer, \eqref{eq:product}--\eqref{eq:Lc}, and the population layer, \eqref{eq:betastar}---and the prior correction restores the dispersion that the naive product understates. The aggregate is formed analytically: \eqref{eq:Lc} is evaluated in closed form and the resulting Gaussian is sampled to provide the item's consensus draws.

\subsection{Eligibility and the population fallback}
\label{sec:fallback}

The subtraction in \eqref{eq:Lc} can render $\bLam_\rmc$ non-positive-definite, as tends to happen for items that the period data barely identify beyond the population model or whose Gaussian summaries are noisy. Rather than regularize $\bLam_\rmc$ (for example by pseudo-inversion), which would impute dispersion the data do not support, we flag these items and assign them the population prediction $\bxi_j\sim\Norm(\bX_j\bbeta^\star,\bSig^\star)$---the replacement prior of \eqref{eq:Lc}. The proportion of items routed to the fallback is a diagnostic of whether the periods are individually informative enough for the procedure.

\section{Empirical Evaluation}
\label{sec:eval}

\subsection{Data and configuration}
\label{sec:data}

The procedure was applied to operational response data from the Duolingo English Test, a high-volume, continuously evolving language assessment \citep{nydick2026maintaining,naismith2025techmanual}, measuring two latent dimensions with items in several response formats (selected-response, partial-credit, and continuous). Each period was calibrated by a scalable Bayesian explanatory IRT engine \citep{nydick2026spice} with period-specific item posteriors thinned to approximately $1{,}000$ draws. The run comprised four consecutive quarterly calibration periods spanning approximately one year with the third period defining the reference metric. Each period included on the order of tens of thousands of items and hundreds of thousands of examinees. Linking used the robust per-draw criterion of Section~\ref{sec:linking} with the $\widehat{R}$-and-purification anchor screen, solved for $200$ draws of each non-reference period; the transformed draws form that period's linked sample. Aggregation used the analytic consensus method of Section~\ref{sec:consensus} with the prior correction: each period's Gaussian summary was estimated from its linked draws, and the consensus Gaussian was sampled to provide $2{,}000$ draws per item.

\subsection{Validation protocol}
\label{sec:protocol}

To obtain evidence that the aggregate reproduces a single pooled analysis, we fit one here as a benchmark. To avoid confounding the results, the benchmark is fit with a separate ability distribution per period---a multiple-group model in which each period's examinees form a group---matching the ability structure of the per-period calibrations.

All comparisons are restricted to converged item parameters: an (item, parameter) pair enters only if its split-chain $\widehat{R}$ does not exceed $1.1$ in the pooled run and in every period run containing the item, since posterior summaries that have not converged on at least one side of the comparison are not interpretable. This screen removed $1.9\%$ of (item, parameter) pairs.

Three comparisons are reported. (i)~\emph{Recovery of item parameters}: the correlation and root-mean-square error (RMSE) of the posterior means, and the same statistics for the posterior standard deviations (Figure~\ref{fig:meansd}; Table~\ref{tab:agreement}). Agreement of the means indicates unbiased linking and aggregation; agreement of the standard deviations indicates that the posterior dispersion, and not only its location, is recovered. (ii)~\emph{Uncertainty by exposure}: because prior over-counting most affects items observed sparsely within each period, the ratio of posterior standard deviations, $\mathrm{SD}_{\mathrm{cons}}/\mathrm{SD}_{\mathrm{pool}}$, is examined within tertiles of per-period response volume (Table~\ref{tab:agreement}, SD-ratio rows); the signature of over-counting is a ratio below one at low exposure that approaches one as exposure increases, whereas a ratio above one indicates residual linking variability. (iii)~\emph{Ability-distribution agreement}: because the item link and the ability-metric transformation share the same constants, comparing each period's ability mean and standard deviation checks the linking independently of the items.

\subsection{Results}
\label{sec:results}

\begin{figure}[t]
  \centering
  \includegraphics[width=\columnwidth]{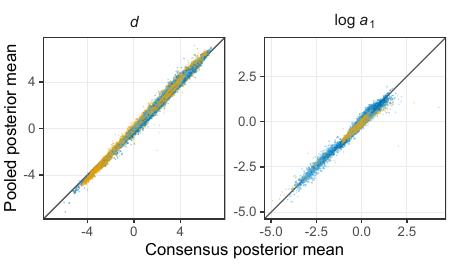}\\[2pt]
  \includegraphics[width=\columnwidth]{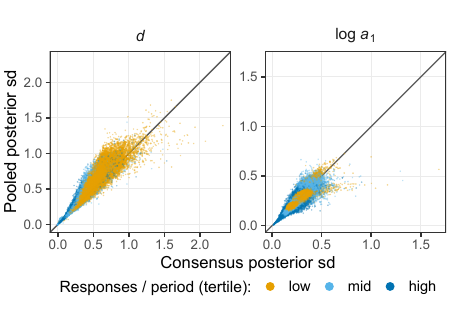}
  \caption{Item-level agreement between the consensus aggregate and the pooled
  benchmark, for item parameters that converged in the pooled run and in every
  period run ($\widehat{R}\le 1.1$). \emph{Top:} posterior means;
  \emph{bottom:} posterior standard deviations; columns are item parameters;
  the gray line is the identity. Color gives per-period exposure tertile.}
  \label{fig:meansd}
\end{figure}

\begin{table}[t]
  \centering
  \small
  \begin{tabular}{llcc}
\toprule
 & & $d$ & $\log a_1$ \\
\midrule
Post.\ mean & $r$ & 0.998 & 0.991 \\
 & RMSE & 0.129 & 0.081 \\
\addlinespace
Post.\ SD & $r$ & 0.970 & 0.920 \\
 & RMSE & 0.066 & 0.036 \\
\addlinespace
SD ratio & low expo. & 0.91 & 0.96 \\
 & mid expo. & 0.94 & 0.95 \\
 & high expo. & 0.94 & 0.98 \\
\bottomrule
\end{tabular}

  \caption{Agreement of the consensus aggregate with the pooled benchmark, for
  converged item parameters only: correlation $r$ and RMSE between the
  consensus and pooled posterior means and posterior standard deviations, and
  the mean posterior-SD ratio (consensus/pooled) within low/mid/high tertiles
  of per-period exposure. A ratio below one indicates residual
  under-dispersion from prior over-counting.}
  \label{tab:agreement}
\end{table}

Figure~\ref{fig:meansd} plots the consensus against the pooled posterior, item by item, for the posterior mean (top) and the posterior standard deviation (bottom), with points colored by per-period exposure tertile; Table~\ref{tab:agreement} summarizes the agreement. Posterior means agreed closely ($r=.998$ for $d$ and $.991$ for $\log a_1$; RMSE $0.129$ and $0.081$), indicating that the per-draw linking and the precision-weighted aggregation introduce no systematic distortion of location. Posterior standard deviations also agreed ($r=.970$ and $.920$; RMSE $0.066$ and $0.036$), evidence that the prior correction recovers most of the dispersion of the pooled posterior and not only its location; the SD-ratio rows quantify the remaining shortfall.

The exposure diagnostic of comparison~(ii) appears in the SD-ratio rows of Table~\ref{tab:agreement}: $0.91/0.94/0.94$ for $d$ and $0.96/0.95/0.98$ for $\log a_1$ across low, mid, and high per-period exposure. Both parameters show mild under-dispersion in the direction expected of residual prior over-counting, particularly in the lowest-exposure tertile.

For comparison~(iii), the per-period ability means implied by the linked period calibrations differed by at most $0.08$ reference-metric units, and the ratio of the corresponding ability standard deviations ranged from $0.98$ to $1.07$.

\section{Discussion}
\label{sec:discussion}

Consensus calibration is an instance of computational psychometrics, bringing scalable Bayesian computation to hierarchical IRT calibration (\hyperlink{cite.vondavier2022compsych}{A.~A. von Davier et al.}, \citeyear{vondavier2022compsych}). It reconstructs the pooled posterior from independent per-period fits. The periods are linked by a robust characteristic-curve criterion solved separately for each posterior draw, in the manner of \citet{baldwin2011common}, which propagates the uncertainty of the linking transformation. The linked posteriors are then combined, with the prior contributed by each period removed and a single prior obtained by consensus reinstated. The cost of an update consequently scales with the new period rather than the accumulated history, and the full history need not be held in memory at once. In the benchmark comparison, the consensus posterior means and standard deviations agreed closely with those of the pooled fit.

The procedure builds on feature-based parameter prediction rather than replacing it: the population model aggregated at its second layer is the regression of item parameters on item features that such prediction estimates, so calibration refines those predictions rather than discarding them. Because the link is solved per draw, the linked posteriors carry the linking error, together with its dependence on the item parameters, forward as calibrated uncertainty \citep{nydick2026maintaining,jewsbury2025variance}.

The procedure has limitations. Cross-period conditional independence requires disjoint examinee samples, since substantial overlap would count some responses more than once. The factors combined are plug-in Gaussian summaries of each period's hierarchical posterior, an approximation whose validity was assessed empirically in the benchmark comparison. Finally, the aggregation targets each item's marginal posterior; the joint dependence across items induced by shared hyperparameters and common linking constants is not reconstructed.

\bibliography{custom}

@article{settles2020machine,
  author  = {Settles, Burr and LaFlair, Geoffrey T. and Hagiwara, Masato},
  title   = {Machine Learning--Driven Language Assessment},
  journal = {Transactions of the Association for Computational Linguistics},
  year    = {2020},
  volume  = {8},
  pages   = {247--263},
  doi     = {10.1162/tacl_a_00310}
}

@inproceedings{mccarthy2021jumpstarting,
  author    = {McCarthy, Arya D. and Yancey, Kevin P. and LaFlair, Geoffrey T.
               and Egbert, Jesse and Liao, Manqian and Settles, Burr},
  title     = {Jump-Starting Item Parameters for Adaptive Language Tests},
  booktitle = {Proceedings of the 2021 Conference on Empirical Methods in
               Natural Language Processing (EMNLP)},
  year      = {2021},
  pages     = {883--899},
  publisher = {Association for Computational Linguistics},
  address   = {Online and Punta Cana, Dominican Republic},
  doi       = {10.18653/v1/2021.emnlp-main.67}
}

@article{vondavier2018aig,
  author  = {von Davier, Matthias},
  title   = {Automated Item Generation with Recurrent Neural Networks},
  journal = {Psychometrika},
  year    = {2018},
  volume  = {83},
  number  = {4},
  pages   = {847--857},
  doi     = {10.1007/s11336-018-9608-y}
}

@misc{sharpnack2024autoirt,
  author       = {Sharpnack, James and Mulcaire, Phoebe and Bicknell, Klinton and
                  LaFlair, Geoffrey T. and Yancey, Kevin P.},
  title        = {{AutoIRT}: Calibrating Item Response Theory Models with Automated
                  Machine Learning},
  year         = {2024},
  howpublished = {arXiv:2409.08823 [cs.LG]},
  doi          = {10.48550/arXiv.2409.08823}
}

@article{benedetto2023survey,
  author  = {Benedetto, Luca and Cremonesi, Paolo and Caines, Andrew and
             Buttery, Paula and Cappelli, Andrea and Giussani, Andrea and
             Turrin, Roberto},
  title   = {A Survey on Recent Approaches to Question Difficulty Estimation
             from Text},
  journal = {ACM Computing Surveys},
  year    = {2023},
  volume  = {55},
  number  = {9},
  pages   = {178:1--178:37},
  doi     = {10.1145/3556538}
}

@article{alkhuzaey2024review,
  author  = {AlKhuzaey, Samah and Grasso, Floriana and Payne, Terry R. and
             Tamma, Valentina},
  title   = {Text-Based Question Difficulty Prediction: A Systematic Review of
             Automatic Approaches},
  journal = {International Journal of Artificial Intelligence in Education},
  year    = {2024},
  volume  = {34},
  pages   = {862--914},
  doi     = {10.1007/s40593-023-00362-1}
}

@inproceedings{yaneva2024bea,
  author    = {Yaneva, Victoria and North, Kai and Baldwin, Peter and Ha, Le An and
               Rezayi, Saed and Zhou, Yiyun and Ray Choudhury, Sagnik and
               Harik, Polina and Clauser, Brian},
  title     = {Findings from the First Shared Task on Automated Prediction of
               Difficulty and Response Time for Multiple-Choice Questions},
  booktitle = {Proceedings of the 19th Workshop on Innovative Use of NLP for Building
               Educational Applications (BEA 2024)},
  year      = {2024},
  pages     = {470--482},
  publisher = {Association for Computational Linguistics},
  address   = {Mexico City, Mexico}
}

@book{gierl2013aig,
  editor    = {Gierl, Mark J. and Haladyna, Thomas M.},
  title     = {Automatic Item Generation: Theory and Practice},
  year      = {2013},
  publisher = {Routledge},
  address   = {New York}
}

@incollection{vondavier2024itemfactory,
  author    = {von Davier, Alina A. and Runge, Andrew and Park, Yena and
               Attali, Yigal and Church, Jacqueline and LaFlair, Geoffrey T.},
  title     = {The Item Factory: Intelligent Automation in Support of Test
               Development at Scale},
  booktitle = {Machine Learning, Natural Language Processing, and Psychometrics},
  editor    = {Jiao, Hong and Lissitz, Robert W.},
  publisher = {Information Age Publishing},
  address   = {Charlotte, NC},
  year      = {2024},
  pages     = {1--25}
}

@book{fox2010bayesian,
  author    = {Fox, Jean-Paul},
  title     = {Bayesian Item Response Modeling: Theory and Applications},
  year      = {2010},
  publisher = {Springer},
  address   = {New York}
}

@book{deboeck2004explanatory,
  editor    = {De Boeck, Paul and Wilson, Mark},
  title     = {Explanatory Item Response Models: A Generalized Linear and
               Nonlinear Approach},
  year      = {2004},
  publisher = {Springer},
  address   = {New York}
}

@article{karabatsos2009bayesian,
  author  = {Karabatsos, George and Walker, Stephen G.},
  title   = {A {Bayesian} Nonparametric Approach to Test Equating},
  journal = {Psychometrika},
  year    = {2009},
  volume  = {74},
  number  = {2},
  pages   = {211--232},
  doi     = {10.1007/s11336-008-9096-6},
}

@book{vondavier2022compsych,
  editor    = {von Davier, Alina A. and Mislevy, Robert J. and Hao, Jiangang},
  title     = {Computational Psychometrics: New Methodologies for a New Generation of Digital Learning and Assessment, with Examples in {R} and {Python}},
  year      = {2022},
  publisher = {Springer},
  address   = {Cham, Switzerland},
  series    = {Methodology of Educational Measurement and Assessment},
  doi       = {10.1007/978-3-030-74394-9},
}

@article{haebara1980,
  author  = {Haebara, Tomokazu},
  title   = {Equating Logistic Ability Scales by a Weighted Least Squares Method},
  journal = {Japanese Psychological Research},
  year    = {1980},
  volume  = {22},
  number  = {3},
  pages   = {144--149}
}

@article{stockinglord1983,
  author  = {Stocking, Martha L. and Lord, Frederic M.},
  title   = {Developing a Common Metric in Item Response Theory},
  journal = {Applied Psychological Measurement},
  year    = {1983},
  volume  = {7},
  number  = {2},
  pages   = {201--210}
}

@book{kolen2014equating,
  author    = {Kolen, Michael J. and Brennan, Robert L.},
  title     = {Test Equating, Scaling, and Linking: Methods and Practices},
  edition   = {3rd},
  year      = {2014},
  publisher = {Springer},
  address   = {New York}
}

@article{candell1988iterative,
  author  = {Candell, Gregory L. and Drasgow, Fritz},
  title   = {An Iterative Procedure for Linking Metrics and Assessing Item Bias
             in Item Response Theory},
  journal = {Applied Psychological Measurement},
  year    = {1988},
  volume  = {12},
  number  = {3},
  pages   = {253--260},
  doi     = {10.1177/014662168801200304}
}

@manual{robitzsch2024sirt,
  author = {Robitzsch, Alexander},
  title  = {sirt: Supplementary Item Response Theory Models},
  year   = {2024},
  url    = {https://CRAN.R-project.org/package=sirt},
  note   = {R package; implements robust Haebara linking with $L_p$ loss}
}

@article{vondavier2007unified,
  author    = {von Davier, Matthias and von Davier, Alina A.},
  title     = {A Unified Approach to {IRT} Scale Linking and Scale Transformations},
  journal   = {Methodology},
  year      = {2007},
  volume    = {3},
  number    = {3},
  pages     = {115--124},
  doi       = {10.1027/1614-2241.3.3.115},
  publisher = {Hogrefe},
}

@article{jewsbury2025variance,
  author  = {Jewsbury, Paul A.},
  title   = {Generally Applicable Variance Estimation Methods for
             Common-Population Linking},
  journal = {Journal of Educational and Behavioral Statistics},
  year    = {2025},
  volume  = {50},
  number  = {4},
  pages   = {651--681},
  doi     = {10.3102/10769986241263976}
}

@article{jewsbury2026achievement,
  author  = {Jewsbury, Paul A.},
  title   = {Linking Error on Achievement Levels Accounting for Dependencies
             and Complex Sampling},
  journal = {Journal of Educational Measurement},
  year    = {2026},
  volume  = {63},
  number  = {1},
  pages   = {e12439},
  doi     = {10.1111/jedm.12439}
}

@article{jewsbury2025invariance,
  author  = {Jewsbury, Paul A.},
  title   = {Standard Error Estimation for Subpopulation Non-invariance},
  journal = {Applied Psychological Measurement},
  year    = {2025},
  volume  = {49},
  number  = {8},
  pages   = {477--505},
  doi     = {10.1177/01466216251351947}
}

@article{baldwin2011common,
  author  = {Baldwin, Peter},
  title   = {A Strategy for Developing a Common Metric in {Item} {Response}
             {Theory} When Parameter Posterior Distributions Are Known},
  journal = {Journal of Educational Measurement},
  year    = {2011},
  volume  = {48},
  number  = {1},
  pages   = {1--11},
  doi     = {10.1111/j.1745-3984.2010.00127.x}
}

@article{scott2016consensus,
  author  = {Scott, Steven L. and Blocker, Alexander W. and Bonassi, Fernando V.
             and Chipman, Hugh A. and George, Edward I. and McCulloch, Robert E.},
  title   = {{Bayes} and Big Data: The Consensus {Monte Carlo} Algorithm},
  journal = {International Journal of Management Science and Engineering Management},
  year    = {2016},
  volume  = {11},
  number  = {2},
  pages   = {78--88},
  doi     = {10.1080/17509653.2016.1142191}
}

@inproceedings{neiswanger2014embarrassingly,
  author    = {Neiswanger, Willie and Wang, Chong and Xing, Eric P.},
  title     = {Asymptotically Exact, Embarrassingly Parallel {MCMC}},
  booktitle = {Proceedings of the Thirtieth Conference on Uncertainty in
               Artificial Intelligence (UAI)},
  year      = {2014},
  pages     = {623--632},
  publisher = {AUAI Press},
  address   = {Arlington, VA}
}

@article{hinton2002poe,
  author  = {Hinton, Geoffrey E.},
  title   = {Training Products of Experts by Minimizing Contrastive Divergence},
  journal = {Neural Computation},
  year    = {2002},
  volume  = {14},
  number  = {8},
  pages   = {1771--1800}
}

@article{tresp2000bcm,
  author  = {Tresp, Volker},
  title   = {A {Bayesian} Committee Machine},
  journal = {Neural Computation},
  year    = {2000},
  volume  = {12},
  number  = {11},
  pages   = {2719--2741}
}

@article{pinheiro1996unconstrained,
  author  = {Pinheiro, Jos\'{e} C. and Bates, Douglas M.},
  title   = {Unconstrained Parametrizations for Variance-Covariance Matrices},
  journal = {Statistics and Computing},
  year    = {1996},
  volume  = {6},
  number  = {3},
  pages   = {289--296}
}

@article{lin2019cholesky,
  author  = {Lin, Zhenhua},
  title   = {{Riemannian} Geometry of Symmetric Positive Definite Matrices via
             {Cholesky} Decomposition},
  journal = {SIAM Journal on Matrix Analysis and Applications},
  year    = {2019},
  volume  = {40},
  number  = {4},
  pages   = {1353--1370},
  doi     = {10.1137/18M1221084}
}

@article{gelman1992rhat,
  author  = {Gelman, Andrew and Rubin, Donald B.},
  title   = {Inference from Iterative Simulation Using Multiple Sequences},
  journal = {Statistical Science},
  year    = {1992},
  volume  = {7},
  number  = {4},
  pages   = {457--472}
}

@article{robitzsch2023invariance,
  author  = {Robitzsch, Alexander and L\"{u}dtke, Oliver},
  title   = {Why Full, Partial, or Approximate Measurement Invariance Are Not
             a Prerequisite for Meaningful and Valid Group Comparisons},
  journal = {Structural Equation Modeling: A Multidisciplinary Journal},
  year    = {2023},
  volume  = {30},
  number  = {6},
  pages   = {859--870},
  doi     = {10.1080/10705511.2023.2191292}
}

@incollection{nydick2026maintaining,
  author    = {Nydick, Steven W. and Liao, Manqian and Chen, Siyuan and Lockwood, J. R.},
  title     = {Maintaining the Meaning of Test Scores in a High-Volume Evolving Assessment},
  booktitle = {The {Routledge} Handbook of Digital Language Assessment: Innovations
               and Insights from the {Duolingo English Test}},
  editor    = {Naismith, Ben and von Davier, Alina A. and Burstein, Jill and LaFlair, Geoffrey T.},
  publisher = {Routledge},
  year      = {2026},
  note      = {In press}
}

@misc{nydick2026spice,
  author       = {Nydick, Steven W. and Liao, Manqian and Lockwood, J. R.},
  title        = {A Scalable Parametric Item Calibration Engine ({SPICE}) for Explanatory
                  {IRT} with Sparse Data},
  year         = {2026},
  howpublished = {arXiv:2605.21782 [stat.ME]},
  doi          = {10.48550/arXiv.2605.21782}
}

@article{xu2024divide,
  author  = {Xu, Sainan and Lu, Jing and Zhang, Jiwei and Wang, Chun and Xu, Gongjun},
  title   = {Optimizing Large-Scale Educational Assessment with a ``Divide-and-Conquer''
             Strategy: Fast and Efficient Distributed {Bayesian} Inference in {IRT} Models},
  journal = {Psychometrika},
  year    = {2024},
  volume  = {89},
  number  = {4},
  pages   = {1119--1147},
  doi     = {10.1007/s11336-024-09978-1}
}

@incollection{sharpnack2026,
  author    = {Sharpnack, James and Nydick, Steven W. and Lockwood, J. R. and Tsigler, Alexander
               and von Davier, Alina A.},
  title     = {Calibration, Scoring, and Administration of a High-Stakes Computerized
               Adaptive Test},
  booktitle = {The {Routledge} Handbook of Digital Language Assessment: Innovations
               and Insights from the {Duolingo English Test}},
  editor    = {Naismith, Ben and von Davier, Alina A. and Burstein, Jill and LaFlair, Geoffrey T.},
  publisher = {Routledge},
  year      = {2026},
  note      = {In press}
}

@inproceedings{yancey2024bertirt,
  author    = {Yancey, Kevin P. and Runge, Andrew and LaFlair, Geoffrey T. and Mulcaire, Phoebe},
  title     = {{BERT-IRT}: Accelerating Item Piloting with {BERT} Embeddings and
               Explainable {IRT} Models},
  booktitle = {Proceedings of the 19th Workshop on Innovative Use of NLP for Building
               Educational Applications (BEA 2024)},
  year      = {2024},
  pages     = {428--438},
  publisher = {Association for Computational Linguistics},
  address   = {Mexico City, Mexico}
}

@techreport{naismith2025techmanual,
  author      = {Naismith, Ben and Cardwell, Ramsey and LaFlair, Geoffrey T. and
                 Nydick, Steven W. and Kostromitina, Masha},
  title       = {{Duolingo English Test}: Technical Manual},
  institution = {Duolingo},
  type        = {Duolingo research report},
  year        = {2025},
  url         = {https://go.duolingo.com/dettechnicalmanual}
}

\appendix

\section{Derivation of the prior correction}
\label{app:deriv}

All densities are over the item parameter vector $\bxi$ on the reference metric, and each period enters through the Gaussian summaries of Section~\ref{sec:consensus}: its linked item posterior $p_m$ and prior contribution $\pi_m$ are the Gaussian summaries, for which the algebra below is exact; relative to the hierarchical posterior, in which the prior is estimated and the periods share hyperparameters, the result is a plug-in approximation. The target posterior combines the period likelihoods with a single prior $\pi^\star$, whereas the product of period posteriors carries the product of $M$ priors; dividing out the per-period priors and multiplying in $\pi^\star$ gives
\[
p^\star(\bxi) \;\propto\; \Big[\textstyle\prod_m p_m(\bxi)\Big]\,
\frac{\pi^\star(\bxi)}{\prod_m \pi_m(\bxi)} .
\]
A Gaussian with mean $\mathbf{a}$ and precision $\mathbf{P}$ contributes $-\tfrac12\bxi^{\!\top}\mathbf{P}\bxi+\bxi^{\!\top}\mathbf{P}\mathbf{a}$ to the log density up to a constant. With period posteriors of precision $\bLam_m$ and mean $\hat\bmu_m$, per-period priors of precision $\bOmega_m$ and mean $\bmm_m$, and replacement prior $(\bOmega^\star,\bmm^\star)$, collecting quadratic and linear terms yields $\bLam_\rmc=\sum_m\bLam_m-\sum_m\bOmega_m+\bOmega^\star$ and $\bb_\rmc=\sum_m\bLam_m\hat\bmu_m-\sum_m\bOmega_m\bmm_m+\bOmega^\star\bmm^\star$, with $\bmu_\rmc=\bLam_\rmc^{-1}\bb_\rmc$, valid whenever $\bLam_\rmc$ is positive definite. Each quotient $p_m/\pi_m$ is an unnormalized Gaussian factor whose precision $\bLam_m-\bOmega_m$ need not itself be positive definite; only $\bLam_\rmc$ must be. Positive-definiteness is the eligibility condition of Section~\ref{sec:fallback}; the population fallback is used when it fails. Equal fixed priors recover $\bLam_\rmc=\sum_m\bLam_m-(M-1)\bOmega$, the combination rule of the Bayesian committee machine \citep{tresp2000bcm}; the same multiplicity of the prior is handled in consensus Monte Carlo and embarrassingly parallel MCMC by assigning each subset the prior raised to the power $1/M$ \citep{scott2016consensus,neiswanger2014embarrassingly}. The correction above generalizes these devices to priors that are estimated, hierarchical, and different across periods, and to a replacement prior obtained by consensus.

\section{Log-Cholesky mean of covariances}
\label{app:logchol}

Let a positive-definite $\bSig$ have lower Cholesky factor $\mathbf{L}$ with positive diagonal. Its log-Cholesky coordinates are the logarithm of the diagonal together with the strict lower-triangular entries of $\mathbf{L}$, a smooth bijection onto an unconstrained Euclidean space whose inverse exponentiates the diagonal and returns $\mathbf{L}\mathbf{L}^{\!\top}$, necessarily positive definite \citep{pinheiro1996unconstrained}. A weighted arithmetic mean taken in these coordinates, with nonnegative weights summing to one, is the weighted Fr\'{e}chet mean under the log-Cholesky metric on covariance matrices \citep{lin2019cholesky}: it matches the unconstrained parameterization in which the covariances are estimated, is positive definite by construction, and has determinant equal to the weighted geometric mean of the period determinants, averaging dispersion on a multiplicative scale and avoiding the determinant inflation of the entrywise mean. Period $m$ is weighted by its normalized item count in the block, and the coordinate ordering is held fixed across periods.

\end{document}